\documentclass{emsprocart}
\usepackage{dsfont}
\usepackage[latin1]{inputenc}
\usepackage{mathrsfs}
\usepackage{cite}
\contact[mathieu.lewin@math.cnrs.fr]{Mathieu Lewin, CNRS \& University of Cergy-Pontoise, Department of Mathematics, UMR CNRS 8088, F-95 000 Cergy-Pontoise Cedex, France}

\newtheorem{theorem}{Theorem}[section]

\theoremstyle{definition}

\newtheorem{remark}[theorem]{Remark}

\newcommand\R{{\ensuremath {\mathbb R} }}
\newcommand\C{{\ensuremath {\mathbb C} }}
\newcommand\1{{\ensuremath {\mathds 1} }}
\renewcommand\phi{\varphi}
\newcommand{\alp}{\boldsymbol{\alpha}}
\newcommand{\alphaph}{\alpha_{\rm ph}}

\newcommand{\gH}{\mathfrak{H}}

\newcommand{\bA}{\boldsymbol{A}}

\renewcommand{\to}{\rightarrow}

\newcommand{\cE}{\mathcal{E}}
\newcommand{\cC}{\mathcal{C}}

\newcommand\ii{{\ensuremath {\infty}}}
\newcommand\pscal[1]{{\ensuremath{\left\langle #1 \right\rangle}}}

\newcommand{\tr}{\operatorname{Tr}}
\newcommand{\Hdiv}{\dot{H}^1_{\rm div}(\R^3)}

\title[A nonlinear variational problem in relativistic quantum mechanics]{A nonlinear variational problem in relativistic quantum mechanics}

\author[Mathieu Lewin]{Mathieu Lewin\thanks{Grants from the French Ministry of Research (ANR BLAN-10-0101) and from the European Research Council under the European Community's Seventh Framework Programme (FP7/2007-2013 Grant Agreement MNIQS 258023) are gratefully acknowledged.}}

\begin{document}

\begin{abstract}
We describe several recent results obtained in collaboration with P.~Gravejat, C.~Hainzl, \'E.~Séré and J.P.~Solovej, 
concerning a nonlinear model for the relativistic quantum vacuum in interaction with a classical electromagnetic field. 
\end{abstract}

\begin{classification}
Primary 35Q40; Secondary 81V10.
\end{classification}

\begin{keywords}
Nonlinear analysis, variational methods, quantum mechanics, Dirac operator, renormalization, quantum electrodynamics, vacuum polarization

\bigskip

\noindent\textsl{Proceedings of the 6th European Congress of Mathematics, Krakow (Poland), July 2012.}
\end{keywords}

\maketitle

%%%%%%%%%%%%%%%%%%%%%%%%%%%%%%%%%
%%%%%%%%%%%%%%%%%%%%%%%%%%%%%%%%%
\section{Introduction}
%%%%%%%%%%%%%%%%%%%%%%%%%%%%%%%%%
%%%%%%%%%%%%%%%%%%%%%%%%%%%%%%%%%

\emph{Quantum Electrodynamics} (QED) is one of the most successful and precise theory in physics. Already studied in the early years of quantum mechanics by Dirac, Pauli, Heisenberg, Fermi, Weisskopf and others, it was finally formulated in its definite form by Bethe, Dyson, Feynman, Schwinger and Tomonaga between 1946 and 1950. This gave Feynman, Schwinger and Tomonaga a Nobel prize in 1965.

The theory is the combination of quantum mechanics and Einstein's special relativity. It aims at describing the interactions between matter and light, at the microscopic scale where quantum effects are dominant. It allows to determine the time-dependent behavior of charged particles (like the \emph{electrons} in an atom or a molecule) when they are coupled to \emph{photons} (the quanta of light). 
Quantum Electrodynamics has an important symmetry called \emph{charge conjugation}. The latter means that the behavior of positively charged and negatively charged particles is described in a similar manner. More importantly, the theory predicts that any charged particle automatically has an anti-particle with the same mass but an opposite charge (for the electron, this particle is called the \emph{positron}), and that it is possible to create a particle/anti-particle pair by providing a sufficient amount of energy to the vacuum. From Einstein's famous relation, this energy must be at least $2\times mc^2$.
Because one can create matter from energy, the vacuum cannot be seen anymore as an empty and inert object as it is considered in everyday life. At the microscopic scale, the quantum vacuum is a fluctuating complicated system which participates to any physical phenomenon. 

Quantum Electrodynamics is an extremely accurate theory. Its predictions are the most precise ever obtained from a physical model, when compared with experiments. The most famous successful predictions are the Lamb shift in the spectrum of the hydrogen atom and the electron anomalous magnetic dipole moment. The agreement with experiment is within a window of about $10^{-8}$.

In more technical terms, Quantum Electrodynamics is an \emph{abelian gauge field theory}. The ideas of Dyson, Feynman, Schwinger and Tomonaga were later used to invent more complicated \emph{non-abelian} gauge field (a.k.a.~Yang-Mills) theories, like those describing the strong and weak forces. 
In units where the speed of light is $c=1$ and Plank's constant is $\hbar=1$, QED only depends on two parameters whose value has to be determined by experiment: the mass $m$ of the electron, and the Sommerfeld coupling constant $\alpha$ which is the square of the electron charge, $\alpha=e^2$. Of course, there might also be external fields which are applied to the system. The physical value of the coupling constant $\alpha$ is small (about $1/137$) and, in the physical literature, Quantum Electrodynamics is always formulated as a \emph{perturbative theory}. This means that the interesting physical quantities are formally expanded as a power series in $\alpha$, and that only the coefficients of the series are computed explicitly, order by order. But there are divergences occurring and these coefficients are all infinite! In order to solve this problem, a regularization parameter $\Lambda$ has to be introduced in the model. Then, the divergences are absorbed by making a change of variable for $m$ and $\alpha$, in order to obtain a well-defined formal series in the limit $\Lambda\to\ii$. This procedure is called \emph{renormalization}.

It is fair to say that this perturbative formulation of QED is perfectly rigorous in the sense that there is a unique and well defined way to get the final answer (which, in most cases, is then in a surprising agreement with experiment). But, on the other hand, the perturbative nature is certainly frustrating from a mathematical point of view. In a famous quotation, Feynman himself said in 1985:
\begin{quotation}
The shell game that we play (...) is technically called `renormalization'. But no matter how clever the word, it is still what I would call a dippy process! Having to resort to such hocus-pocus has prevented us from proving that the theory of quantum electrodynamics is mathematically self-consistent. It's surprising that the theory still hasn't been proved self-consistent one way or the other by now; I suspect that renormalization is not mathematically legitimate.~\cite{Feynman-90}
\end{quotation}
There has been no dramatic change since Feynman's quotation, and renormalization has become a common (and somehow accepted) tool.

%\medskip

On the mathematical side, there have been several works on models originating from QED, but not so many on the true theory itself. We should recall that one of the famous Millenium Prize Problems of the Clay Mathematics Institute concerns the construction of a well-defined Yang-Mills theory as well as the understanding of its low energy excitation spectrum. Yang-Mills theory gathers all non-abelian gauge field theories as alluded above, and any deeper understanding of the (abelian) theory of Quantum Electrodynamics is therefore desirable.

In the last few years, mathematicians have been particularly interested in studying the interaction of light with \emph{non relativistic} matter. In this simplified model the electrons are quantum but they are assumed to have a speed which is much smaller than the speed of light, such that relativistic effects can be neglected. Their description then involves the Laplacian instead of the Dirac operator. Events like the creation of electron-positron pairs are encoded in the Dirac operator and they cannot be described within such a non-relativistic theory. But it is already a very important and fundamental problem to understand the effect of quantized light on non-relativistic particles. Some of the recent works concern for instance the existence of a lower eigenvalue for the underlying Hamiltonian~\cite{BacFroSig-99,Hiroshima-99,GriLieLos-01}, resonances and the relaxation to the ground state~\cite{BacFroSig-98,BacFroSig-99,AboFauFroSig-09}, problems related to divergences and mass renormalization~\cite{HaiSei-02b,LieLos-02,BacCheFroSig-07,BacFroPiz-09,HasHer-11}, and the stability of large systems~\cite{LieLosSol-95,FefFroGra-97,Lieb-03,LieLos-05}.

Coming back to relativistic particles, some authors have considered Lattice QED~\cite{OstSei-78, Seiler-82}. Other studied linear and nonlinear models based on the Dirac operator for finitely many particles in an `inert' vacuum and with a classical electromagnetic field, see e.g.~\cite{Georgiev-91,EstGeoSer-96,EstSer-99,FlaSimTaf-97,HubSie-07,DolEstLos-07} and the references in~\cite{EstLewSer-08}. Finally, the quantum vacuum and the process of pair creation was investigated in a non-interacting setting (meaning without any light at all), e.g. in~\cite{KlaSch-77b,KlaSch-77a,NenSch-78,Nenciu-87,PicDur-08}.

With P.~Gravejat, C.~Hainzl, \'E.~Séré and J.-P.~Solovej, we followed another route in a series of works~\cite{HaiLewSer-05a, HaiLewSer-05b, HaiLewSol-07, HaiLewSerSol-07, HaiLewSer-09, GraLewSer-09, GraLewSer-11,GraHaiLewSer-12} which originated from a fundamental paper of Chaix and Iracane~\cite{ChaIra-89,ChaIraLio-89}, and which was stimulated by the previous works~\cite{BacBarHelSie-99,HaiSie-03}. We considered relativistic particles in a fluctuating vacuum, both described by the Dirac equation, and in interaction with a \emph{classical electromagnetic field}. That light is not quantized in this model prevents us from describing important physical effects. But, on the other hand, these seem to be the first mathematical results dealing with the quantum vacuum in interaction with light. Also, as we will explain, we are able to construct the associated model in a fully non-perturbative fashion, in a simplified setting.

The purpose of this article is to review these recent results. We will particularly insist on the uncommon mathematical aspects, like those related to renormalization, and which are related to important physical effects. The reader interested in knowing more details can also read the last section of the review~\cite{EstLewSer-08}.

%%%%%%%%%%%%%%%%%%%%%%%%%%%%%%%%%
%%%%%%%%%%%%%%%%%%%%%%%%%%%%%%%%%
\section{The quantum vacuum in classical electromagnetic fields}
%%%%%%%%%%%%%%%%%%%%%%%%%%%%%%%%%
%%%%%%%%%%%%%%%%%%%%%%%%%%%%%%%%%

In this section we present a model for quantum relativistic particles evolving in a fluctuating quantum vacuum, and interacting with a classical electromagnetic field. We will not discuss too much here the mathematical meaning of the equations, which will be explained in the next sections. Actually, most of the terms that we write here are infinite quantities if no special care is employed. We will concentrate on the time-independent model and look for stationary states. 

Our system is composed of particles (electrons or positrons) together with the quantum vacuum on the one side, and of a classical electromagnetic field $(E,B)$ describing light on the other side, all evolving in the physical space $\R^3$. It is useful to use electromagnetic potentials $V$ and $A$ which are such that $E=-\nabla V$ and $B=\nabla\wedge A$. We might also consider fixed external fields $E_{\rm ext}=-\nabla V_{\rm ext}$ and $B_{\rm ext}=\nabla\wedge A_{\rm ext}$ which are applied to the system.
The particles and the vacuum are described by the Dirac operator
\begin{equation}
D_{m,e\bA_{\rm tot}}:=\sum_{k=1}^3\alpha_k\Big(-i\frac{\partial}{\partial{x_k}}-eA_{\rm tot}(x)_k\Big)+m\beta-eV_{\rm tot}(x),
\label{eq:Dirac_operator}
\end{equation}
where $\bA_{\rm tot}:=(V_{\rm tot},A_{\rm tot})$ with $V_{\rm tot}=V+V_{\rm ext}$ and $A_{\rm tot}=A+A_{\rm ext}$
are the total electromagnetic fields, and $e$ is the (bare) electron charge.
The four Dirac matrices $\alp = (\alpha_1, \alpha_2 , \alpha_3)$ and $\beta$ are equal to
$$\alpha_k := \begin{pmatrix} 0 & \sigma_k \\ \sigma_k & 0 \end{pmatrix} \quad {\rm and} \quad \beta := \begin{pmatrix} I_2 & 0 \\ 0 & - I_2 \end{pmatrix},$$
the Pauli matrices $\sigma_1$, $\sigma_2$ and $\sigma_3$ being defined by
$$\sigma_1 := \begin{pmatrix} 0 & 1 \\ 1 & 0 \end{pmatrix},\quad \sigma_2 := \begin{pmatrix} 0 & - i \\ i & 0 \end{pmatrix} \quad {\rm and} \quad \sigma_3 := \begin{pmatrix} 1 & 0 \\ 0 & - 1 \end{pmatrix}.$$
These matrices are chosen to ensure that 
$\big(D_{m,0,0}\big)^2=-\Delta+m^2$
which is the quantum equivalent of Einstein's fundamental relation $E^2=c^2p^2+m^2c^4$ for the relativistic classical energy $E$ in terms of the momentum $p$,
with $c=1$ in our case.
The operator in~\eqref{eq:Dirac_operator} acts on the Hilbert space $\gH:=L^2(\R^3,\C^4)$. Under reasonable assumptions on $V_{\rm tot}$ and $A_{\rm tot}$, it is self-adjoint on the Sobolev space $H^1(\R^3,\C^4)$, see~\cite{Thaller}. The \emph{charge conjugation} is the anti-unitary operator defined on $\gH$ by $\mathscr{C}f:=i\beta\alpha_2\overline{f}$, and which satisfies
%\begin{equation}
$\mathscr{C} D_{m,e\bA_{\rm tot}}\mathscr{C}^{-1}=-D_{m,-e\bA_{\rm tot}}$.
% \label{eq:charge-conjugation}
% \end{equation}
This relation implies that the spectrum of $D_{m,0,0}$ is symmetric with respect to $0$,
$$\sigma\big(D_{m,0,0}\big)=(-\ii,-m]\cup[m,\ii).$$
When $V_{\rm tot}$ and $A_{\rm tot}$ decay at infinity, the essential spectrum stays the same by the Rellich-Kato theorem, $\sigma_{\rm ess}\big(D_{m,e\bA_{\rm tot}}\big)=(-\ii,-m]\cup[m,\ii)$. Isolated eigenvalues of finite multiplicity can however appear in the gap $(-m,m)$ when $V_{\rm tot}\neq0$.

The state of one particle is described by a normalized wave function $\phi\in L^2(\R^3,\C^4)$, $\int_{\R^3}|\phi|^2=1$, and its corresponding energy is $\langle\phi,D_{m,e\bA_{\rm tot}}\phi\rangle$ (which has to be understood in the form sense when $\phi\in H^{1/2}(\R^3,\C^4)$). The fact that the Dirac operator is unbounded from below, hence that the energy can be arbitrarily negative was very surprising for his inventor, P.A.M. Dirac~\cite{Dirac-28,Dirac-28b}. In quantum mechanics it is always assumed that the most stable state of the system is the one with the lowest energy, and there does not seem to be any here. From a mathematical point of view this makes a huge difference as compared to non-relativistic models based on the Laplace operator $-\Delta$. That the energy is unbounded from below means that minimization methods cannot be employed and that one has to resort to complicated min-max techniques to construct solutions~\cite{EstLewSer-08}.

Dirac did not want to renounce the physical picture that states with lower energy are more stable. So, in 1930 he suggested to reinterpret the problem by changing the role of the vacuum:
\begin{quotation}
We make the assumption that, in the world as we know it, nearly all the states of negative energy for the electrons are occupied, with just one electron in each state, and that a uniform filling of all the negative-energy states is completely unobservable to us. \cite{Dirac-34b} 
\end{quotation}
Physically, one therefore has to imagine that the vacuum (called the \emph{Dirac sea}) is filled with infinitely many virtual particles occupying the negative energy states. With this conjecture, a real electron cannot be in a negative state. This is because electrons are \emph{fermions} and that two such particles can never be in the same quantum state (\emph{Pauli principle}).
With this interpretation, Dirac was able to conjecture the existence of ``holes" in the vacuum, interpreted as \emph{anti-electrons} or \emph{positrons}, and which were later experimentally discovered by Anderson \cite{Anderson-33}. Dirac also predicted the phenomenon of vacuum polarization: In an external electric field, the virtual electrons are displaced, and the vacuum acquires a non constant density of charge. The idea of a fluctuating quantum vacuum was born.

Let us now explain Dirac's idea in more mathematical terms. The state of $N$ electrons can be represented by $N$ wave functions $\phi_1,...,\phi_N\in H^1(\R^3,\C^4)$, such that $\pscal{\phi_i,\phi_j}_{L^2}=\delta_{ij}$, the latter constraint being the mathematical formulation of the Pauli principle. The total energy is the sum of the energies of the individual electrons, which can be written as
$$\sum_{j=1}^N\pscal{\phi_j,D_{m,e\bA_{\rm tot}}\phi_j}=\tr \big(D_{m,e\bA_{\rm tot}} P\big),$$
where $P$ is the orthogonal projector onto the $N$-dimensional space spanned by the $\phi_j$'s. Using the bra-ket notation we have $P=\sum_{j=1}^N|\phi_j\rangle\langle\phi_j|$. Now if we allow any number $N$ of particles and look for the state of lowest energy, the formal solution is the negative spectral projector
\begin{equation}
P=P^-_{m,e\bA_{\rm tot}}:=\1_{(-\ii,0)}\big(D_{m,e\bA_{\rm tot}}\big).
\label{eq:proj} 
\end{equation}
Think of a finite hermitian matrix $M$, then the solution to the minimization problem $\inf\{\tr(MP),\ P^2=P=P^*\}$ is the negative spectral projector $P=\1_{(-\ii,0)}(M)$ of $M$, and the corresponding energy is $-\tr M_-$ with $x_-:=-\min(x,0)$ denoting the negative part. The minimizer is unique when $\ker(M)=\{0\}$.
In our case the state in~\eqref{eq:proj} corresponds to filling with particles all the negative energies of $D_{m,e\bA_{\rm tot}}$ as suggested by Dirac. This state $P^-_{m,e\bA_{\rm tot}}$ is the quantum vacuum in the presence of the electromagnetic potential $\bA_{\rm tot}$. It has infinitely many particles and the corresponding total energy is also infinite. 

\begin{remark}\label{rmk:trace}
There are several ways to give a mathematical meaning to the assertion that $P^-_{m,e\bA_{\rm tot}}$ minimizes $P\mapsto \tr (D_{m,e\bA_{\rm tot}} P)$. The first is to use a so-called \emph{thermodynamic limit}. The ambient space $\gH=L^2(\R^3,\C^4)$ is approximated by a sequence of finite-dimensional spaces in which everything makes sense and it is proved that in the limit the minimizer converges to $P^-_{m,e\bA_{\rm tot}}$. The second possibility is to directly argue in the whole space $\gH$ that
$$\tr D_{m,e\bA_{\rm tot}} (P-P^-_{m,e\bA_{\rm tot}})\geq0$$
for all projection $P$. Under appropriate assumptions on $P$ (for instance $P-P^-_{m,0}$ finite rank and smooth), we can write
\begin{align}
\tr D_{m,e\bA_{\rm tot}} (P-P^-_{m,e\bA_{\rm tot}})&=\tr |D_{m,e\bA_{\rm tot}}|\Big(P^+_{m,e\bA_{\rm tot}}(P-P^-_{m,e\bA_{\rm tot}})P^+_{m,e\bA_{\rm tot}}\nonumber\\
&\quad - P^-_{m,e\bA_{\rm tot}}(P-P^-_{m,e\bA_{\rm tot}})P^-_{m,e\bA_{\rm tot}}\Big)\nonumber\\
&=\tr |D_{m,e\bA_{\rm tot}}|^{1/2}\Big(P^+_{m,e\bA_{\rm tot}}PP^+_{m,e\bA_{\rm tot}}\nonumber\\
&\quad + P^-_{m,e\bA_{\rm tot}}(1-P)P^-_{m,e\bA_{\rm tot}}\Big)|D_{m,e\bA_{\rm tot}}|^{1/2}\geq0.\label{eq:def_gen_trace}
\end{align}
We have used here the commutativity of the trace and that $D_{m,e\bA_{\rm tot}}=|D_{m,e\bA_{\rm tot}}|$ $\big(P^+_{m,e\bA_{\rm tot}}-P^-_{m,e\bA_{\rm tot}}\big)$, by definition of the spectral projections. The last term in~\eqref{eq:def_gen_trace} is the trace of a non-negative operator which is always well-defined in $[0,\ii]$, and which can be taken as a definition for the trace on the left side. See~\cite[Sec. 2.1]{HaiLewSer-05a} for a systematic and abstract theory of traces of this form.
\end{remark}

When $\bA_{\rm tot}=0$, then $P^-_{m,0}$ is nothing but the negative spectral projector of the free Dirac operator $D_{m,0}$, which represents the \emph{free vacuum}. This operator is translation invariant and, even if the corresponding charge density is infinite, it is somehow constant. This is the `uniformity' alluded to in Dirac's quotation. When $e\bA_{\rm tot}\neq0$, the state of the vacuum changes. Later we will want to optimize over $\bA$ with $\bA_{\rm ext}$ fixed, but before we have to modify a bit our theory in order to make it charge-conjugation invariant. 
If we take an arbitrary state $P=P^2$, then we get 
$\tr \big(D_{m,e\bA_{\rm tot}} P\big)=-\tr \big(D_{m,-e\bA_{\rm tot}} \mathscr{C}P\mathscr{C}^{-1}\big)$.
This is not very satisfactory because $-\mathscr{C}P\mathscr{C}^{-1}$ is not a fermionic state and we cannot reinterpret this as the energy of something. It is better to subtract half the identity to the operator $P$, which amounts to adding a (infinite) constant. So we consider instead the energy
$$\tr D_{m,e\bA_{\rm tot}} \left(P-\frac12\right)=\tr D_{m,e\bA_{\rm tot}} \frac{P-P^\perp}2$$
with $P^\perp:=1-P$. Now we see that
$\tr D_{m,e\bA_{\rm tot}} (P-P^\perp)=\tr D_{m,-e\bA_{\rm tot}}(P'-{P'}^\perp)$ with $P':=\mathscr{C}P^\perp\mathscr{C}^{-1}$.
The subtraction of half the identity to $P$ is a common technique for systems at half filling like ours. It is explained using the formalism of second quantization in~\cite{HaiLewSol-07}.

We are now able to write the total Lagrangian of our system, which is the (formal) sum of the particle energy and Maxwell's classical Lagrangian: 
$$\mathscr{L}_{m,e}(P,\bA\;;\bA_{\rm ext}):=\tr D_{m,e(\bA+\bA_{\rm ext})} \left(P-\frac12\right)+\frac{1}{8\pi}\int_{\R^3}|B|^2-|E|^2,$$
where we recall that $E=-\nabla V$ and $B=\nabla\wedge A$. Our purpose is to look for critical points of this Lagrangian, obtained by minimizing over $P$ and $A$ and maximizing over $V$, as is usually done in classical electrodynamics. We have already explained that the minimum over $P$ with $\bA$ fixed gives $P=P^-_{m,e\bA_{\rm tot}}$. It is also possible to optimize over $\bA$, with $P$ fixed. The optimal potentials solve Gauss' equation
$$-\Delta V=4\pi\,e\,\rho_{P-1/2},\qquad -\Delta A=4\pi\,e\,j_{P-1/2}$$
where $\rho_{P-1/2}$ and $j_{P-1/2}$ are the (formal) density of charge and density of current of the vacuum in the state $P$. These are defined by $\rho_M(x)=\tr_{\C^4} M(x,x)$ and $j_M(x)_k=\tr_{\C^4} \alpha_kM(x,x)$,
for any locally trace-class operator $M$ with integral kernel $M(x,y)$.
Note that these are zero for the free vacuum:
$$\rho_{P^-_{m,0}-1/2}\equiv0,\qquad j_{P^-_{m,0}-1/2}\equiv0,$$
as is shown using the commutation properties of the matrices $\alpha_k$. This emphasizes the importance of subtracting half the identity to our vacuum state.

Our task is to optimize over both $P$ and $\bA$ and it is reasonable to think that the free vacuum $P=P^-_{m,0}$ is the optimal state when $\bA_{\rm ext}=0$. But, the functional $\mathscr{L}_{m,e}$ is infinite and so we cannot easily optimize it. We can however subtract the universal (infinite) constant $\mathscr{L}_{m,e}(P^-_{m,0},0\;;0)$ and consider the relative Lagrangian
\begin{multline}
\mathscr{L}_{m,e}^{\rm rel}(P,\bA\;;\bA_{\rm ext})=\tr D_{m,0}\big(P-P^-_{m,0}\big)-e\int_{\R^3}j_{P-P^-_{m,0}}\cdot (A+A_{\rm ext})\\-e\int_{\R^3}\rho_{P-P^-_{m,0}}(V+V_{\rm ext})+\frac{1}{8\pi}\int_{\R^3}|B|^2-|E|^2.
\label{eq:relative_Lagrangian}
\end{multline}
Because the difference $P-P^-_{m,0}$ is a much better behaved operator than $P-1/2$, we will be able to give a clear meaning to this Lagrangian. The purely electrostatic case $A=A_{\rm ext}=0$ is much easier to deal with than the general case, and we discuss it first in the next section.

%%%%%%%%%%%%%%%%%%%%%%%%%%%%%%%%%
%%%%%%%%%%%%%%%%%%%%%%%%%%%%%%%%%
\section{The purely electrostatic case}
%%%%%%%%%%%%%%%%%%%%%%%%%%%%%%%%%
%%%%%%%%%%%%%%%%%%%%%%%%%%%%%%%%%

%%%%%%%%%%%%%%%%%%%%%%%%%%%%%%%%%
%\subsection{Introducing a Fourier cut-off}
%%%%%%%%%%%%%%%%%%%%%%%%%%%%%%%%%
In the relative Lagrangian~\eqref{eq:relative_Lagrangian}, we now assume $A=A_{\rm ext}=0$ and we maximize with respect to $V$. We find the following energy functional
$$\tr D_{m,0}Q-e\int_{\R^3}\rho_{Q}V_{\rm ext}+\frac\alpha2 D(\rho_Q,\rho_Q)$$
with $\alpha=e^2$ (the Sommerfeld coupling constant), $Q:=P-P^-_{m,0}$ and 
$$D(f,g):=\int_{\R^3}\int_{\R^3}\frac{\overline{f(x)}\,g(y)}{|x-y|}dx\,dy=4\pi\int_{\R^3}\frac{\overline{\hat{f}(k)}\,\hat{g}(k)}{|k|^2}\,dk$$
which is the Coulomb (i.e. $\dot{H}^{-1}(\R^3)$) scalar product. The associated space will be denoted by
$\cC:=\{f\in \mathscr{S}'(\R^3)\ :\ D(f,f)<\ii\}$.
In our setting we are interested in taking for $V_{\rm ext}$ the external field induced by some localized density of charge $\nu_{\rm ext}$, describing for instance the nuclei of a molecule. So we take $V_{\rm ext}=e\nu_{\rm ext}\ast|x|^{-1}$ and get the energy
\begin{equation}
\cE_{m,\alpha}^{\nu_{\rm ext}}(Q)=\tr D_{m,0}Q-\alpha D(\rho_Q,\nu_{\rm ext})+\frac\alpha2 D(\rho_Q,\rho_Q)
\label{eq:energy} 
\end{equation}
which is called the reduced Bogoliubov-Dirac-Fock energy~\cite{ChaIra-89,HaiLewSer-05b}. 

Under suitable assumptions on $\nu_{\rm ext}$, the minimization of this functional in $Q$ makes perfect sense. First we remark that, by the Cauchy-Schwarz inequality,
$$-\alpha D(\rho_Q,\nu_{\rm ext})+\frac\alpha2 D(\rho_Q,\rho_Q)\geq -\frac{\alpha}{2}D(\nu_{\rm ext},\nu_{\rm ext})$$
which is finite when $\nu_{\rm ext}\in\cC$ (for instance for $\nu_{\rm ext}\in L^{6/5}(\R^3)$, by the Hardy-Littlewood-Sobolev inequality~\cite{LieLos-01}). Then we recall that $\tr D_{m,0}Q\geq0$ for any operator $Q=P-P^-_{m,0}$, provided the trace is defined like in~\eqref{eq:def_gen_trace}. This is reminiscent of the fact that $P^-_{m,0}$ formally minimizes $P\mapsto \tr D_{m,0} P$. From all this we conclude that $\cE^{\nu_{\rm ext}}(Q)\geq -\alpha D(\nu_{\rm ext},\nu_{\rm ext})/2$ and the energy is bounded from below. 

However we are unlucky that the infimum of $\cE^{\nu_{\rm ext}}_{m,\alpha}$ is \emph{never attained}~\cite[Thm 2]{HaiLewSer-05b}, except when $\nu_{\rm ext}\equiv0$ where $Q=0$ is the optimizer. The clear mathematical statement is
\begin{equation*}
\inf_{-P^-_{m,0}\leq Q\leq P^+_{m,0}}\left(\tr D_{m,0}Q-\alpha D(\rho_Q,\nu)+\frac\alpha2 D(\rho_Q,\rho_Q)\right)\\=-\frac{\alpha}{2}D(\nu,\nu).
\end{equation*}
In the infimum one can restrict to finite-rank smooth operators $Q$, which ensures that all the terms make sense.
We see that there is never a minimizer when $\nu_{\rm ext}\neq0$, because it should satisfy both $\rho_Q=\nu$ and $Q=0$. As explained in~\cite{HaiLewSer-05b}, this issue is due to certain divergences in Fourier space at infinity. 

In order to give a meaning to our minimization problem, we have to impose a cut-off for the large Fourier frequencies. There are several ways to do this, some being more natural than others. In the purely electric case, we can consider a simple sharp cut-off consisting in replacing our ambient Hilbert space $\gH$ by
$$\gH_\Lambda:=\left\{f\in L^2(\R^3,\C^4)\ :\ \widehat{f}(k)=0 \text{ for } |k|\geq\Lambda\right\}.$$
Note that $D_{m,0}$ maps $\gH_\Lambda$ into itself. For simplicity we use the same notation $D_{m,0}$ and $P^\pm_{m,0}$ for the associated restrictions to $\gH_\Lambda$. The operator $Q$ is defined on $\gH_\Lambda$ and its energy has the same expression as in~\eqref{eq:energy}. It now has minimizers.

\begin{theorem}[Existence of the polarized vacuum~\cite{HaiLewSer-05a,HaiLewSer-05b}]\label{thm:existence}
Let $m>0$, $0<\Lambda<\ii$ and $\alpha\geq0$. For any fixed $\nu_{\rm ext}\in \cC=\dot{H}^{-1}(\R^3)$, the minimization problem
\begin{equation}
\inf\Big\{\cE^{\nu_{\rm ext}}_{m,\alpha}(Q)\ :\ Q:\gH_\Lambda\to\gH_\Lambda,\ -P^-_{m,0}\leq Q\leq P^+_{m,0},\ \mp \tr P^\pm_{m,0}Q P^\pm_{m,0}<\ii\Big\} 
\label{eq:inf_rBDF}
\end{equation}
admits at least one minimizer $Q_*$. All these minimizers share the same density $\rho_{Q_*}$. Any  minimizer $Q_*$ is a solution of the nonlinear equation
\begin{equation}
\begin{cases}
Q_*=\1_{(-\ii,0)}\left(D_*\right)-P^-_{m,0}+\delta\\[0,2cm]
D_*=\Pi_\Lambda\Big(D_{m,0}+\alpha(\rho_{Q_*}-\nu_{\rm ext})\ast|x|^{-1}\Big)\Pi_\Lambda
\end{cases}
\label{eq:SCF}
\end{equation}
where $\Pi_\Lambda$ is the projection onto $\gH_\Lambda$, and $0\leq \delta\leq \1_{\{0\}}\left(D_*\right)$.

If $\alpha\pi^{1/6}2^{11/6}D(\nu,\nu)^{1/2}<\sqrt{m}$, then $\ker\left(D_*\right)=\{0\}$ and $\tr(P^+_{m,0}Q_* P^+_{m,0}+P^-_{m,0}Q_* P^-_{m,0})=0$. Hence $\delta\equiv0$ in~\eqref{eq:SCF}, and the minimizer $Q_*$ is unique. 
\end{theorem}

The equation~\eqref{eq:SCF} on the infinite rank operator $P_*=Q_*+P^-_{m,0}$ can be interpreted as an infinite system of coupled nonlinear partial differential equations. As we have said we always use the convention 
$$\tr D_{m,0} Q:=\tr |D_{m,0}|\big(Q^{++}-Q^{--}\big)$$
where $Q^{++}:=P^+_{m,0}QP^+_{m,0}\geq0$ and $Q^{--}:=P^-_{m,0}QP^-_{m,0}\leq0$ when $Q$ satisfies the constraint $-P^-_{m,0}\leq Q\leq P^+_{m,0}$. Since with the cut-off $\Lambda$ the operator $D_{m,0}$ is bounded, it is sufficient to ask that
$Q^{++}$ and $Q^{--}$ are trace-class in order to properly define the trace, as required in~\eqref{eq:inf_rBDF}. On the other hand simple algebraic manipulations show that the constraint is equivalent to
\begin{equation}
-P^-_{m,0}\leq Q\leq P^+_{m,0} \Longleftrightarrow Q^2\leq Q^{++}-Q^{--}
\label{eq:constraint} 
\end{equation}
as remarked in~\cite{BacBarHelSie-99}. So we see that we must also have $\tr Q^2<\ii$. In other words, for the first term in the energy to be finite, $Q$ must be a Hilbert-Schmidt operator whose diagonal blocks $Q^{\pm\pm}$ are both trace-class. Under these conditions and with the ultraviolet cut-off $\Lambda$, it was proved in~\cite{HaiLewSer-09} that $\rho_Q\in L^2(\R^3)\cap \cC$. This is sufficient to prove Theorem~\ref{thm:existence} using simple arguments from convex analysis. In~\cite{HaiLewSer-05a,HaiLewSer-05b} a more complicated model with an additional so-called exchange term is considered and this term makes the analysis much more involved.

Let us emphasize that Theorem~\ref{thm:existence} is valid for \emph{any value of the coupling constant $\alpha$} and \emph{any value of the ultraviolet cut-off $\Lambda$}. This is therefore a non-perturbative result. Because no smallness assumption is made on the external density $\nu_{\rm ext}$, the model is appropriate for the description of non-perturbative events like the spontaneous creation of electron-positron pairs in strong external fields~\cite{Sabin-11}. Note that for simplicity we have minimized the energy with the relaxed constraint~\eqref{eq:constraint} instead of the original constraint that $P:=Q+P^-_{m,0}=P^2$. However, for a small external density $\nu_{\rm ext}$, the minimizer $Q_*$ is unique and the corresponding state $P_*$ is an orthogonal projection, as expected. 

In Theorem~\ref{thm:existence} we have constructed the polarized quantum vacuum in the presence of the external density $\nu_{\rm ext}$, which is by definition a global minimizer of the energy. It is also possible to construct states having a certain number $N$ of `real' electrons (or positrons). More precisely, we can minimize the energy under a charge constraint of the form $\text{``}\tr Q\text{''}:=\tr(Q^{++}+Q^{--})=N$, see~\cite{HaiLewSer-09,GraLewSer-09}. Minimizers do not always exist, depending on the value of $N$ and on the strength of the density $\nu_{\rm ext}$. Estimates on the maximal number of electrons that can be bound by $\nu_{\rm ext}$ were provided in~\cite{GraLewSer-09}, following ideas of Lieb~\cite{Lieb-84}. Any minimizer with charge constraint  solves a nonlinear equation similar to~\eqref{eq:SCF}, with the energy level $0$ replaced by a Lagrange multiplier $\mu\in(-m,m)$:
\begin{equation}
Q_*=\1_{(-\ii,\mu)}\left(D_*\right)-P^-_{m,0}+\delta
\label{eq:SCF_charge}
\end{equation}
where now $0\leq \delta\leq \1_{\{\mu\}}\left(D_*\right)$. If $\mu>0$, splitting $\1_{(-\ii,\mu)}=\1_{(-\ii,0)}+\1_{(0,\mu)}$ gives the corresponding states for the polarized vacuum and for the electrons. If $\mu<0$, we split $\1_{(-\ii,\mu)}=\1_{(-\ii,0)}-\1_{(\mu,0)}$ and get the state of the polarized vacuum and of the positrons. In the rest of this article we will not discuss further this constrained minimization problem, and we will instead concentrate on the pure vacuum case considered in Theorem~\ref{thm:existence}.

The minimizers obtained in Theorem~\ref{thm:existence} are very singular mathematical objects. The following says that $Q_*$ is never a trace-class operator.

\begin{theorem}[Renormalized charge~\cite{GraLewSer-09}]\label{thm:renormalization}
Let $Q_*$ be a minimizer as in Theorem~\ref{thm:existence}, for some $\nu_{\rm ext}\in L^1(\R^3)\cap \cC$. Then we have $\rho_{Q_*}\in L^1(\R^3)$ and
\begin{equation}
\int_{\R^3}\rho_{Q_*}-\nu_{\rm ext} =\frac{\tr(Q_*^{++}+Q_*^{--})-\displaystyle\int_{\R^3}\nu_{\rm ext}}{1+\alpha B_{\Lambda/m}}
\label{eq:charge_renormalization}
\end{equation}
where
\begin{equation}
B_{\Lambda/m}  = \frac1\pi\int_{0}^{\frac{\Lambda}{\sqrt{m^2+\Lambda^2}}}\frac{z^2-z^4/3}{1-z^2}dz=\frac{2}{3\pi}\log\frac{\Lambda}m-\frac{5}{9\pi}+\frac{2\log2}{3\pi}+O(m^2/\Lambda^2).
\label{expression_B}
\end{equation}
In particular, if $\alpha\pi^{1/6}2^{11/6}D(\nu,\nu)^{1/2}<\sqrt{m}$, then $Q_*$ is not trace-class.
\end{theorem}

Let us recall that a compact operator $B$ is trace-class when $\tr|B|=\tr\sqrt{B^*B}<\ii$. This is equivalent to saying that $\sum_{j}\pscal{\phi_j,|B|\phi_j}<\ii$ for one (hence for all) orthonormal basis $\{\phi_j\}$ of the ambient Hilbert space. By the spectral theorem, this is also the same as saying that the eigenvalues of $B$ are summable, $\sum_j|\lambda_j|<\ii$. If the ambient space is an $L^2$ space, we can write by the spectral theorem $B=\sum_{j}\lambda_j|\psi_j\rangle\langle\psi_j|$ and we get that the corresponding density $\rho_B=\sum_{j}\lambda_j|\psi_j|^2$ is in $L^1$ and that $\int\rho_B=\tr B$.

For a non positive operator $B$, it can happen that $\sum_{j}\pscal{\phi_j,B\phi_j}$ is convergent for one basis and not for another. This is exactly what is happening for our operator $Q_*$. The two diagonal blocks $Q_*^{++}$ and $Q_*^{--}$ are trace-class, which means that $\sum_j\pscal{\phi_j^+,Q_*\phi_j^+}+\pscal{\phi_j^-,Q_*\phi_j^-}$ converges for any orthonormal basis $\{\phi_j^\pm\}$ of $\gH^\pm_\Lambda:=P^\pm_{m,0}\gH_\Lambda$. The surprise is that $\rho_{Q_*}$ is always in $L^1(\R^3)$ but that in general $\int_{\R^3}\rho_{Q_*}\neq \tr(Q^{++}+Q^{--})$. The discrepancy between these two quantities is universal and it is given by the relation~\eqref{eq:charge_renormalization}. The problem comes from the off-diagonal densities $\rho_{Q_*^{\pm\mp}}$ which are in $L^1(\R^3)$ but do not have a vanishing integral. More precisely, only the first order term when $\rho_{Q_*^{\pm\mp}}$ are expanded in a power series in $\alpha$ contribute to $\int_{\R^3}\rho_{Q_*}$. The proof of Theorem~\ref{thm:renormalization} consists in studying this term in details~\cite{GraLewSer-09}.

Theorem~\ref{thm:renormalization} has a natural interpretation in terms of charge renormalization. Imagine that we put a nucleus of charge $Z=\int_{\R^3}\nu_{\rm ext}$ in the vacuum, which is weak in the sense that $D(\nu_{\rm ext},\nu_{\rm ext})$ is small, and let $P_*=Q_*+P^-_{m,0}$ be the corresponding unique polarized vacuum obtained by Theorem~\ref{thm:existence}. This vacuum is neutral,
$\tr(Q_*^{++}+Q_*^{--})=0$, which means that the external field is not strong enough to create electron-positron pairs. In reality we never measure the charge of the nucleus alone, but we always also observe the corresponding vacuum polarization. Hence we do not see $Z$, but rather $Z/(1+\alpha B_{\Lambda/m})$. This corresponds to having a physical coupling constant $\alphaph$ given by the renormalization formula
\begin{equation}
\alpha_{\rm ph}=\frac\alpha{1+\alpha B_{\Lambda/m}} \Longleftrightarrow \alpha=\frac\alphaph{1-\alphaph B_{\Lambda/m}}.
\label{charge_renormalization}
\end{equation}
In our theory we must take $\alpha_{\rm ph}\simeq1/137$, the `bare' $\alpha$ can never been observed. So using the change of variable~\eqref{charge_renormalization} we should express any other physical quantity predicted by our model in terms of $\alphaph$, $m$ and $\Lambda$ only. 

The natural question arises whether it is possible to remove the ultraviolet cut-off $\Lambda$, by keeping $\alphaph$ and $m$ fixed (note that the mass $m$ is not renormalized in this model). The answer is clearly no! We immediately get from~\eqref{charge_renormalization} that $\alphaph B_{\Lambda/m}<1$ and therefore $\alphaph\to0$ whenever we try to take $\Lambda\to\ii$. This phenomenon is called the Landau pole \cite{Landau-55} and one has to look for a weaker definition of renormalizability. The cut-off $\Lambda$ which was first introduced as a mathematical trick to regularize the model has actually a physical meaning. A natural scale occurs beyond which the model does not make sense. Fortunately, this corresponds to momenta of the order $\Lambda\sim me^{3\pi/2\alphaph}$, a huge number for $\alphaph\simeq1/137$.

In~\cite{GraLewSer-11}, the regime $\alphaph\ll1$, $\Lambda\gg1$ with $\alphaph\log\Lambda$ fixed was studied. It was proved that the dressed density of the nucleus admits a Taylor expansion 
\begin{equation}
\alpha(\nu_{\rm ext}-\rho_{Q_*})= \alphaph\nu_{\rm ext}+\sum_{k=1}^K\nu_k\,\alphaph^{k+1}+O(\alphaph^{K+2})
\label{eq:order-by-order} 
\end{equation}
where the terms $\nu_k$ in the expansion are independent of the value of $\alphaph\log\Lambda$. The first correction $\nu_1$ gives rise to the famous Uehling potential~\cite{Uehling-35,Serber-35}. The relation~\eqref{eq:order-by-order} shows that the density of the nucleus can be renormalized order by order. It is believed that the series in~\eqref{eq:order-by-order} is divergent~\cite{Dyson-52}, but no mathematical argument has been provided so far.

To summarize, in the purely electrostatic case we have a well defined non-perturbative theory for all $m>0$, all $\alphaph\geq0$ and all cut-off $\Lambda$ such that $\alphaph B_{\Lambda/m}<1$. In the regime $\alphaph\ll1$, $\Lambda\gg1$ with $\alphaph\log\Lambda$ fixed, the perturbation series computed by the physicists is recovered.

%%%%%%%%%%%%%%%%%%%%%%%%%%%%%%%%%
%%%%%%%%%%%%%%%%%%%%%%%%%%%%%%%%%
\section{The electromagnetic case}
%%%%%%%%%%%%%%%%%%%%%%%%%%%%%%%%%
%%%%%%%%%%%%%%%%%%%%%%%%%%%%%%%%%

Let us go back to our relative Lagrangian~\eqref{eq:relative_Lagrangian}, including now the electromagnetic fields $B_{\rm tot}=B+B_{\rm ext}$. Optimizing with respect to $P$ we find a formal Lagrangian depending only on the classical fields
\begin{equation}
\mathscr{L}_{m,e}^{\rm rel}(\bA\;;\bA_{\rm ext})=\frac12\tr\Big( |D_{m,0}|-|D_{m,e(\bA+\bA_{\rm ext})}|\Big)+\frac{1}{8\pi}\int_{\R^3}|B|^2-|E|^2.
\label{eq:relative_Lagrangian2}
\end{equation}
This Lagrangian is not really well-defined because of the ultraviolet divergences which have to be taken care of. On the contrary to the pure electrostatic case studied in the last section, we cannot impose a sharp cut-off here, as it is extremely important to keep the magnetic gauge invariance corresponding to replacing $A$ by $A+\nabla \phi$. In~\cite{GraHaiLewSer-12} we used the \emph{Pauli-Villars} method~\cite{PauVil-49}, which consists in introducing two fictitious particle fields of very high masses $m_1,m_2\gg1$ which play the role of ultraviolet cut-offs. The Pauli-Villars-regulated Lagrangian is defined by
\begin{equation}
\mathscr{L}_{m,e}^{\rm PV}(\bA\;;\bA_{\rm ext})=\frac12\tr\sum_{j=0}^2c_j\Big( |D_{m_j,0}|-|D_{m_j,e(\bA+\bA_{\rm ext})}|\Big)+\frac{1}{8\pi}\int_{\R^3}|B|^2-|E|^2
\label{eq:relative_Lagrangian_PV}
\end{equation}
where $c_0=1$, $m_0=m$ and the fictitious fields are chosen such that
\begin{equation}
\sum_{j=0}^2c_j=\sum_{j=0}^2c_jm^2_j=0.
\label{cond:PV}
\end{equation}
The main result of~\cite{GraHaiLewSer-12} is the following

\begin{theorem}[The Pauli-Villars-regulated vacuum in electromagnetic fields~\cite{GraHaiLewSer-12}]
Let $e\in\R$, $m>0$ and $c_1,c_2,m_1,m_2$ satisfying~\eqref{cond:PV}.

\smallskip

\noindent $(i)$ The functional 
$\bA\mapsto \tr\;\tr_{\C^4}\sum_{j=0}^2c_j\Big( |D_{m_j,0}|-|D_{m_j,\bA}|\Big)$
is well-defined for $\bA\in C_c^\ii(\R^3,\R^4)$ and it admits a unique continuous extension to
$$\Hdiv=\left\{(V,A)\in L^6(\R^3,\R^4)\ :\ {\rm div} A=0,\ (-\nabla V,\nabla\wedge A)\in L^2(\R^3,\R^6)\right\}.$$
So the Pauli-Villars-regulated Lagrangian~\eqref{eq:relative_Lagrangian_PV} is well-defined and continuous on $\Hdiv\times \Hdiv$.

\medskip

\noindent $(ii)$ For any $\bA_{\rm ext}\in \Hdiv$ with $e \| \bA_{\rm ext} \|_{\Hdiv} < r \sqrt{m}/2$,
there exists a unique solution $\bA_* = (V_*, A_*) \in \Hdiv$ to the min-max problem
% \begin{align*}
% \mathscr{L}_{m,e}^{\rm PV}(\bA_*\;;\bA_{\rm ext}) &= \max_{\| \nabla V \|_{L^2} < \frac{r \sqrt{m}}{e}} \, \mathscr{L}_{m,e}^{\rm PV}(V, A_*\;;\bA_{\rm ext})\\
% & = \min_{\| \nabla\wedge A \|_{L^2} < \frac{r \sqrt{m}}{e}} \, \mathscr{L}_{m,e}^{\rm PV}(V_*, A\;;\bA_{\rm ext}), 
% \end{align*}
% or, equivalently, to 
\begin{equation}
\label{eq:maxmin_minmax}
\begin{split}
\mathscr{L}_{m,e}^{\rm PV}(\bA_*,\bA_{\rm ext}) & = \max_{\| \nabla V \|_{L^2} < \frac{r \sqrt{m}}{e}} \quad \inf_{\| \nabla\wedge A \|_{L^2} < \frac{r \sqrt{m}}{e}} \, \mathscr{L}_{m,e}^{\rm PV}(\bA\;;\bA_{\rm ext})\\
& = \min_{\| \nabla\wedge A \|_{L^2} < \frac{r \sqrt{m}}{e}} \quad \sup_{\| \nabla V \|_{L^2} < \frac{r \sqrt{m}}{e}} \,  \mathscr{L}_{m,e}^{\rm PV}(\bA\;;\bA_{\rm ext}).
\end{split}
\end{equation}
for some radius $r>0$ which only depends on $\sum_{j = 0}^2 |c_j| (m/m_j)$ and provided that $e$ stays in a bounded set.

\medskip

\noindent {\it (ii)} The four-potential $\bA_*=(V_*,A_*)$ is a solution to the nonlinear equations
\begin{equation}
\label{eq:SCF_PV}
\Bigg\{ \begin{array}{ll} - \Delta V_* = 4 \pi e\,\rho_{\bA_*+\bA_{\rm ext}},\\
- \Delta A_* = 4 \pi e\, j_{\bA_*+\bA_{\rm ext}}, \end{array}
\end{equation}
where $\rho_{\bA_* + \bA_{\rm ext}}$ and $j_{\bA_* + \bA_{\rm ext}}$ refer to the charge and current densities of 
\begin{equation}
\label{eq:SCF_op}
Q_*=\sum_{j = 0}^2 c_j \, \1_{(- \infty, 0)} \big( D_{m_j, e (\bA_* + \bA_{\rm ext})} \big).
\end{equation}

\smallskip

\noindent $(iii)$ If $\bA_{\rm ext}\equiv0$, then $\bA_*\equiv0$ as well.
\end{theorem}

This result completely settles the problem to construct Dirac's vacuum in small external fields. It is an open question to do the same in large external fields (which could be useful to understand non-perturbative events like the creation of electron-positron pairs). It is also not known whether our solution is global. For the free vacuum, $\bA_{\rm ext}\equiv\boldsymbol{0}$, we conjecture that the saddle point $\bA\equiv\boldsymbol{0}$ is the unique solution to~\eqref{eq:maxmin_minmax} with $r\equiv\ii$. Even if this was not done in~\cite{GraHaiLewSer-12}, it is possible to renormalize the coupling constant $\alpha$ in the same fashion as in the previous section.

%%%%%%%%%%%%%%%%%%%%%%%%%%%%%%%%%
%%%%%%%%%%%%%%%%%%%%%%%%%%%%%%%%%
%\section{Conclusion and open problems}
%%%%%%%%%%%%%%%%%%%%%%%%%%%%%%%%%
%%%%%%%%%%%%%%%%%%%%%%%%%%%%%%%%%

%%%%%%%%%%%%%%%%%%%%%%%%%%%%%%%
%%%%%%%%%%%%%%%%%%%%%%%%%%%%%%%
% \bibliographystyle{siam}
% \bibliography{biblio.bib}

\end{document}